# The Effect of Electron Lens as Landau Damping Device on Single Particle Dynamics in HL-LHC

A. Valishev, FNAL, Batavia IL 60510, USA

July 25, 2017

*Abstract*
An electron lens can serve as an effective mechanism for suppressing coherent instabilities in high intensity storage rings through nonlinear amplitude dependent betatron tune shift [1]. However, the addition of a strong localized nonlinear focusing element to the accelerator lattice may lead to undesired effects in particle dynamics. We evaluate the effect of a Gaussian electron lens on single particle motion in HL-LHC using numerical tracking simulations, and compare the results to the case when an equal tune spread is generated by conventional octupole magnets.

**Goals**

1. Compare the single particle motion dynamics with a single Gaussian electron lens creating significant nonlinear amplitude-dependent tune shift to the case when an equal tune shift is created with standard lattice octupoles for a realistic scenario of HL-LHC machine with imperfections.
2. Study the robustness of the electron lens scenario against basic machine parameter changes, such as the betatron tune and chromaticity.

**Simulation Parameters**

All simulations were performed with HL-LHC V1.0 lattice (fully squeezed to $\beta^*=0.15$ m) with all multipolar errors included, and nominal beam parameters: number of protons per bunch $N_p=2.2\times10^{11}$, transverse normalized emittance $\varepsilon=2.5$ $\mu$m, bunch length $\sigma_z=7.5$ cm and momentum spread $\sigma_E=0.00011$. The beam-beam interactions were switched off since the action of the Landau Electron Lens is anticipated in the stages of collider cycle preceding the collisions. The betatron tune chromaticity was set to its nominal value of +3 units.

The electron lens element was placed at the candidate location near IR4, where the horizontal and vertical beta-functions are close to being equal ($\beta_x=\beta_y=180$ m, which corresponds to the proton rms beam size of $\sigma_p=0.28$ mm). The electron beam size was matched to the size of the proton beam. The value of electron beam current was chosen to create the betatron tune shift of 0.01 for the particles with amplitude of 3.5 beam sigma.

For the comparison cases where the equal tune spread was created with standard lattice octupoles, the octupole current was –2000 A.

The simulation studies were performed using the macroparticle tracking code *LIFETRAC*. Two types of studies were performed for each simulation case: a) the Frequency Map Analysis on the short time scale of 4096 turns for on-momentum particles; b) the long-term Dynamical Aperture calculation ($10^5$ turns) for particles with momentum deviation of 2.45 $\sigma_E$. This is the standard configuration employed in all LHC and HL-LHC particle tracking campaigns.

**Results**

*1. Nominal tune working point.*

Figure 1 presents the FMA plots in the amplitude space (left) and tune space (right) for the electron lens case in the nominal betatron tune working point of $Q_x=0.31$, $Q_y=0.32$. The color chart on the FMA plots depicts the tune jitter along the particle trajectory in logarithmic scale. Generally, the red color corresponding to jitter of $10^{-3}$ points at either strong resonances or chaotic motion. Blue regious correspond to well-behaved regular particle motion. One can see that in the simulation domain of 8×8 beam sigma, no strong overlapping resonances are observed. The cyan trace on the amplitude plot (left) depicts the long-term Dynamical Aperture boundary for off-momentum particles. The minimum DA for this case is 8 beam sigma (note that the primary collimators are placed at the distance of 6 sigma).

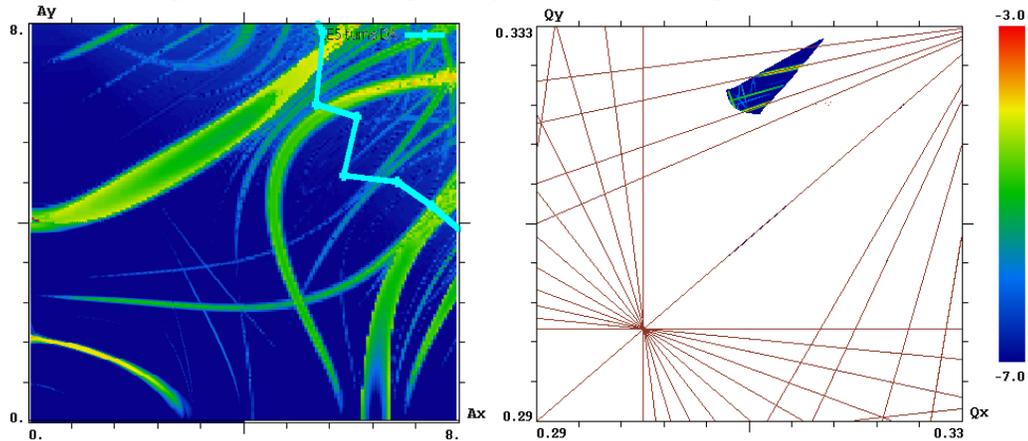

Figure 1. Amplitude space FMA (left) and tune space FMA (right) for Electron Lens and nominal tunes $Q_x=0.31$, $Q_y=0.32$.

Figure 2 shows the results for the octupole case in the same nominal configuration. The much reduced area of stable motion is visible even in the short term FMA simulations (uncolored areas in the amplitude chart depict the lost particles), and is confirmed by the long term tracking study, which yields the Dynamical Aperture of 3.7 sigma.

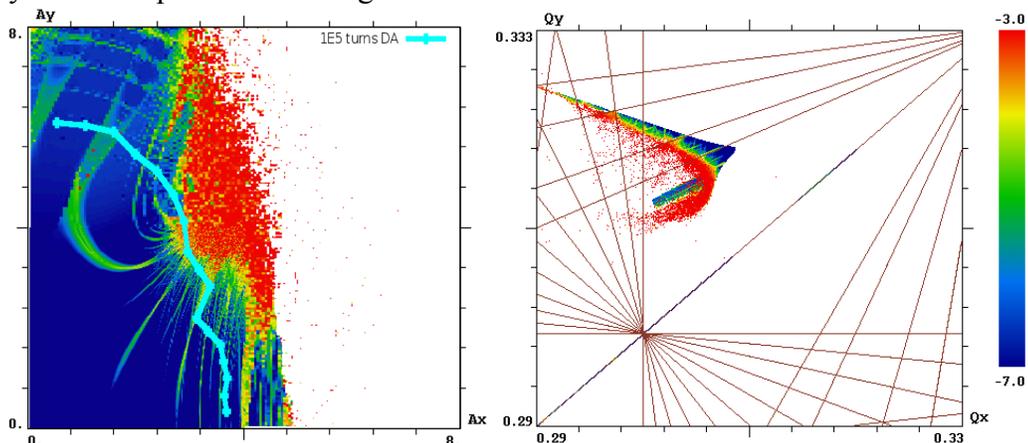

Figure 2. Amplitude space FMA (left) and tune space FMA (right) for octupole case and nominal tunes $Q_x=0.31$, $Q_y=0.32$.

*2. Tune scan.*

In order to study the robustness of the proposed device, a fairly wide betatron tune scan was performed. In this scan, the horizontal and vertical betatron tunes were changed simultaneously by the same amount, i.e. moved along the diagonal. Figures 3, 4 present the results for the electron lens case and the octupole case, respectively, in the betatron tune working point of $Q_x$=0.315, $Q_y$=0.325. Figures 5, 6 show the results for $Q_x$=0.305, $Q_y$=0.315. The plot in Figure 7 summarizes the data for thi tune scan, and shows the minimum DA as a function of the tune change.

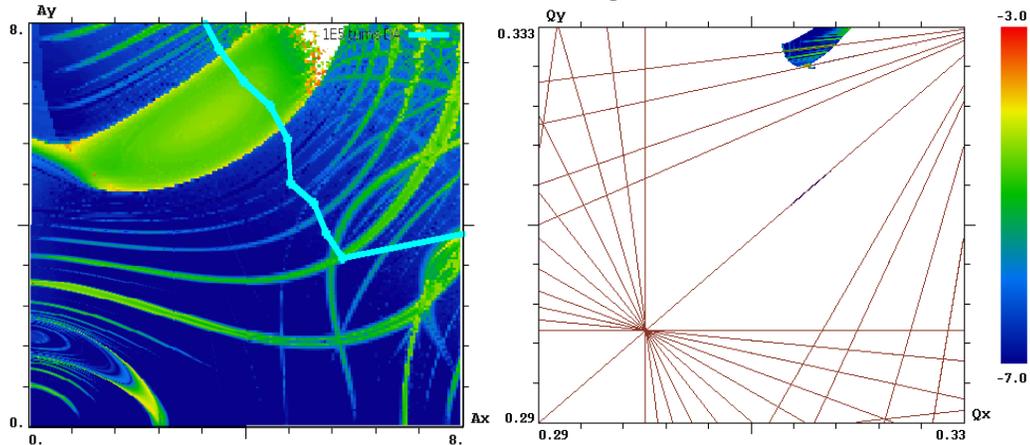

Figure 3. Amplitude space FMA (left) and tune space FMA (right) for Electron Lens case and tunes $Q_x$=0.315, $Q_y$=0.325.

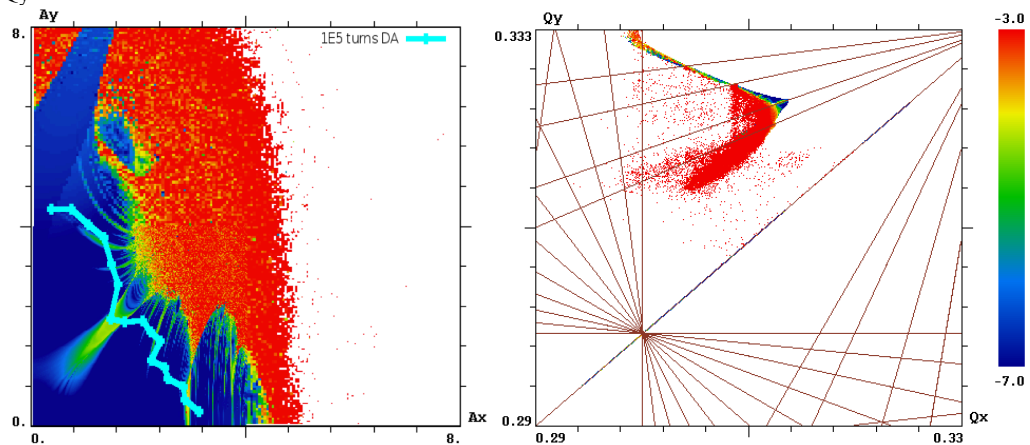

Figure 4. Amplitude space FMA (left) and tune space FMA (right) for octupole case and tunes $Q_x$=0.315, $Q_y$=0.325.

FERMILAB-TM-2659-AD-APC

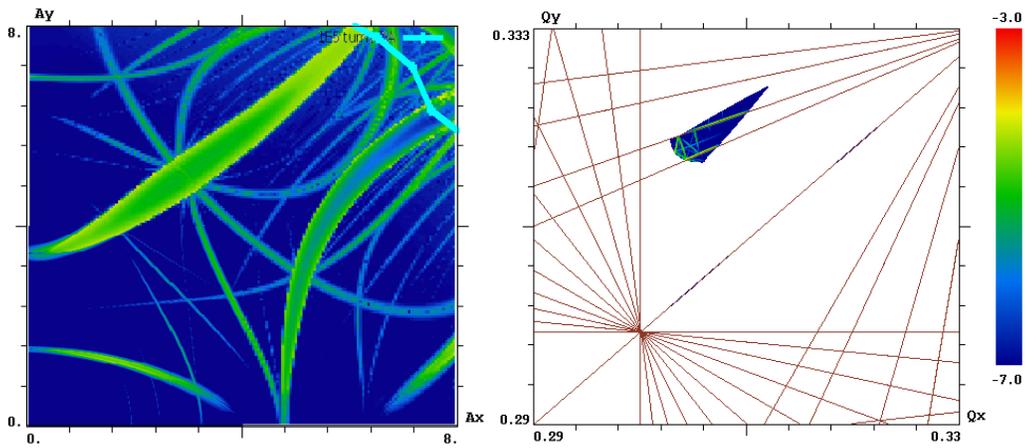

Figure 5. Amplitude space FMA (left) and tune space FMA (right) for Electron Lens case and tunes $Q_x$=0.305, $Q_y$=0.315.

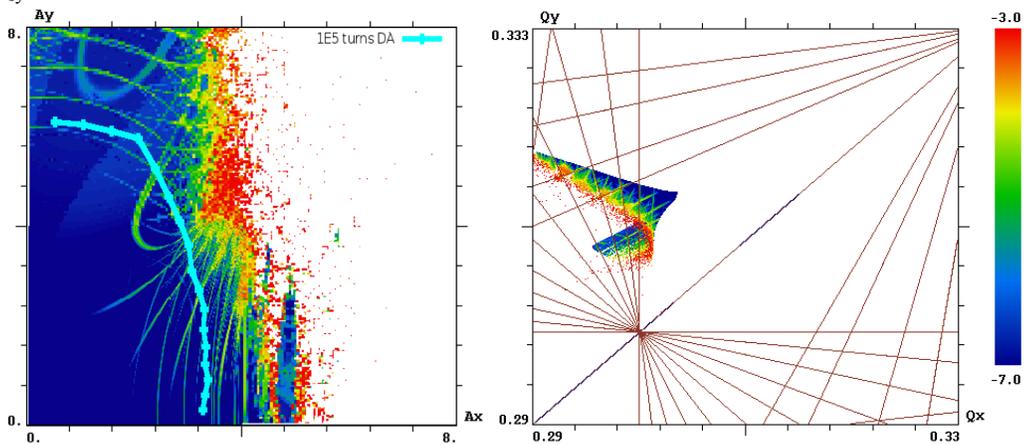

Figure 6. Amplitude space FMA (left) and tune space FMA (right) for octupole case and tunes $Q_x$=0.305, $Q_y$=0.315.

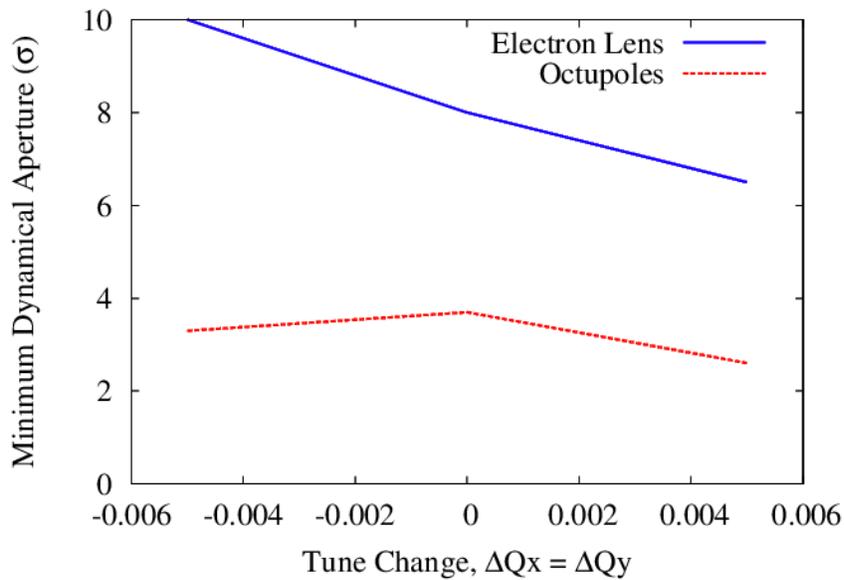

Figure 7. Summary of the tune scan data – minimum Dynamical Aperture as function of the tune change.

## Conclusion

The tracking simulation study of a single Gaussian electron lens generating the nonlinear amplitude dependent tune shift of 0.01, applied to the HL-LHC machine with full set of nonlinear imperfections predicts significant margins in the single-particle motion stability. The solution is robust against the variation of the betatron tune in a fairly wide range of 0.01: the minimum dynamical aperture in all cases was above the safe threshold of 6 beam sigma. In comparison with the Electron Lens, the standard lattice octupoles in the HL-LHC, when required to generate equal nonlinear tune shift, reduce the dynamical aperture to 3 sigma and below, which is unsustainable based on the LHC operational experience.